\documentclass[twoside]{article}
\usepackage{qic}

\usepackage{amsmath}
\usepackage{amssymb}
\usepackage{braket}
\usepackage{color}

\newtheorem{thm}{Theorem}

\newcommand{\miniket}[1]{\vert#1\rangle}

\begin{document}
\setlength{\textheight}{8.0truein}    

\runninghead{A limit law of the return probability for a quantum walk on a hexagonal lattice}
            {T. Machida}

\normalsize\textlineskip
\thispagestyle{empty}
\setcounter{page}{1}


\vspace*{0.88truein}

\alphfootnote

\fpage{1}

\centerline{\bf
A limit law of the return probability for a quantum walk on a hexagonal lattice}
\vspace*{0.37truein}
\centerline{\footnotesize
Takuya Machida}
\vspace*{0.015truein}
\centerline{\footnotesize\it Research Fellow of Japan Society for the Promotion of Science,}
\baselineskip=10pt
\centerline{\footnotesize\it Meiji University, Nakano Campus, 4-21-1 Nakano, Nakano-ku, Tokyo 164-8525, Japan}
\vspace*{0.225truein}

\vspace*{0.21truein}

\abstracts{
A return probability of random walks is one of the interesting subjects.
As it is well known, the return probability strongly depends on the structure of the space where the random waker moves.
On the other hand, the return probability of quantum walks, which are quantum models corresponding to random walks, has also been investigated to some extend lately. 
In this paper, we present a limit of the return probability for a discrete-time 3-state quantum walk on a hexagonal lattice.
}{}{}

\vspace*{10pt}

\vspace*{3pt}

\vspace*{1pt}\textlineskip    

\bibliographystyle{qic}

\section{Introduction}
Quantum walks are quantum analogies of random walks.
As a popular topic in science, random walks have been intensively investigated.
Quantum walks started to get attention in early 2000's because their application to quantum computers was focused on around that time.
We can consider quantum walks as quantum algorithms and applied them to some algorithms~\cite{Venegas-Andraca2012}.
When we solve problems by using classically stochastic algorithms, some of them require exponential number steps for the size of their input to get the solutions.
Since quantum algorithms give quadratically faster speed-up than classical ones, we expect to solve such difficult problems within polynomial number steps by using quantum algorithms.
Quantum walkers on a lattice have coin-states and the behavior of the walkers is described by wave function.
The wave function of a discrete-time quantum walk evolutes according to a unitary operator which is obtained from a product of a coin-flip operator and a position-shift operator.

In this paper we treat a quantum walk on a hexagonal lattice.
Hexagonal lattices show up in various fields of physics.
For example, graphene has a hexagonal structure.
To use quantum walks for applications, it would be important to clarify the behavior of quantum walks on such a realistic lattice.
In particular, we concentrate on computing a limit of a return probability of the quantum walk after long time.
It is well known that the return probability of random walks strongly depends on the structure of a lattice where the random walker moves.
On the other hand, that of the quantum walker depends on its coin-flip operator rather than the structure.
Our walker moves on a 2-dimensional rectangle lattice, but the motion is controlled by position-shift operators and it is equivalent to a motion on a hexagonal lattice.
There are a lot of limit theorems for quantum walks on a line, while a few 2-dimensional walks have been exactly analyzed~\cite{DiMcMachidaBusch2011,WatabeKobayashiKatoriKonno2008,MachidaChandrashekarKonnoBusch2013}.
That's because it is generally much harder to compute 2-dimensional walks than 1-dimensional walks.
Koll{\'a}r et al.~\cite{KollarvStefavnakKissJex2010} treated a 3-state quantum walk on a 2-dimensional plane.
The walker defined by them moves on a hexagon-like lattice and they numerically analyzed the return probability.
The decay order of the return probability of some special 2-dimensional walks were analytically estimated and discussed with the quantum P\'{o}lya number~\cite{vStefavnakKissJex2008,vStefavnakJexKiss2008}.
Though it is a special class of 1-dimensional walks, a relationship between localization of the quantum walks and recurrence properties of the corresponding random walks was clarified~\cite{Segawa2013}.

This paper is organized as follows.
In Sec.~\ref{sec:define}, we define a 3-state quantum walk on a 2-dimensional lattice.
The walker starts from the origin and its motion is controlled by two position-shift operators so that it moves on a hexagonal lattice.
The return probability is analyzed and its long-time limit is calculated in Sec.~\ref{sec:return_probability}.
In the final section we summarize and discuss our result.
Then we will briefly show a limit theorem in a rescaled space by time.

\section{Description of a quantum walk on a hexagonal lattice}
\label{sec:define}
\begin{figure}[h]
 \begin{center}
  \includegraphics[scale=0.5]{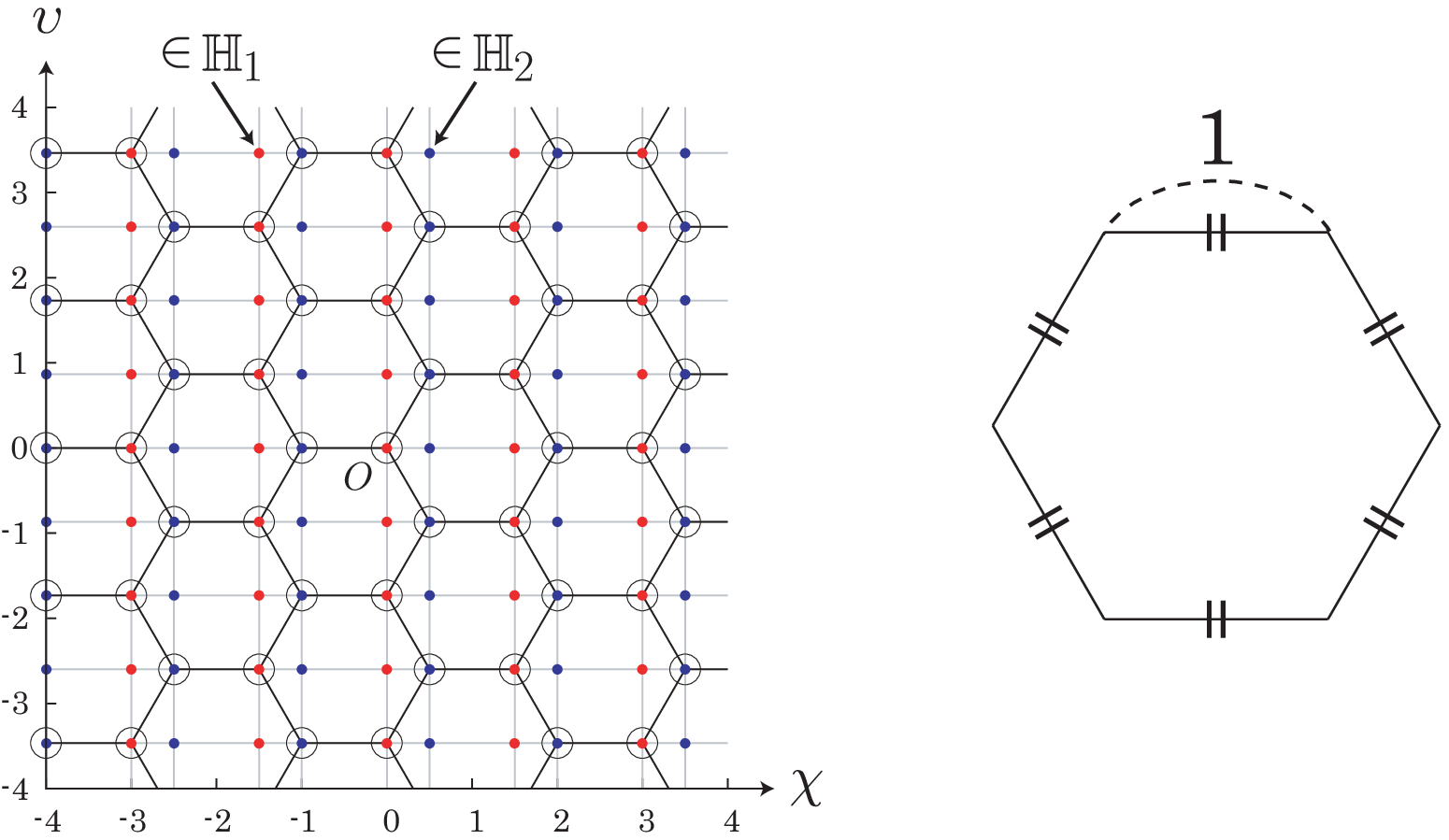}
  \fcaption{Red points are elements in $\mathbb{H}_1$ and blue points are elements in $\mathbb{H}_2$. When the walker starts from the origin $O\,(0,0)$, it moves over the points with a black circle.}
  \label{fig:lattice}
 \end{center}
\end{figure}

First, we define a quantum walk on a two-dimensional lattice $\mathbb{H}=\mathbb{H}_1\cup\mathbb{H}_2$ with $\mathbb{H}_1=\left\{\left(\frac{3x}{2},\frac{\sqrt{3}y}{2}\right) : x,y\in\mathbb{Z}\right\}, \mathbb{H}_2=\left\{\left(\frac{3x+1}{2},\frac{\sqrt{3}y}{2}\right) : x,y\in\mathbb{Z}\right\}$, where $\mathbb{Z}$ means the set of integers.
Then we describe the motion of the walker on a hexagonal lattice by position-shift operators.
Figure~\ref{fig:lattice}, which will show up later, helps us get a picture of both the 2-dimensional lattice $\mathbb{H}$ and the hexagonal lattice which we treat in this paper.
The position of the walker is expressed on two Hilbert spaces $\mathcal{H}_{p_1}=\left\{\ket{\chi,\upsilon} : (\chi,\upsilon)\in\mathbb{H}_1\right\}$ and $\mathcal{H}_{p_2}=\left\{\ket{\chi,\upsilon} : (\chi,\upsilon)\in\mathbb{H}_2\right\}$, and a Hilbert space $\mathcal{H}_p=\mathcal{H}_{p_1}\oplus\mathcal{H}_{p_2}$ just corresponds to the two-dimensional position space where the walker moves.
In addition, the walker at each vertex on $\mathcal{H}_p$ supposes to be in superposition with three coin-states.
We introduce a Hilbert space $\mathcal{H}_c$ which is spanned by the basis $\left\{\ket{0}, \ket{1}, \ket{2}\right\}$ so that the coin space of the walker is expressed. 
To compute a limit law later, we take the following orthonormal vectors: 
\begin{equation}
 \ket{0}=\begin{bmatrix}
	  1\\0\\0
	 \end{bmatrix},\quad
 \ket{1}=\begin{bmatrix}
	  0\\1\\0
	 \end{bmatrix},\quad
 \ket{2}=\begin{bmatrix}
	  0\\0\\1
	 \end{bmatrix}.
\end{equation}
The whole state $\ket{\Psi_t}$ of the quantum walker at time $t\,\in\left\{0,1,2,\ldots\right\}$ is described on the tensor Hilbert space $\mathcal{H}_p\otimes\mathcal{H}_c$.
The position of the walker is shifted by two position-shift operators $S_1, S_2$ after the superposition is operated by a coin-flip operator $C$ as follows:
\begin{equation}
 \miniket{\Psi_{t+1}}=(S_1+S_2)C\ket{\Psi_t},\label{eq:time_evo}
\end{equation}
where
\begin{align}
 S_1=\sum_{x,y\in\mathbb{Z}} & \Biggl\{\Ket{\textstyle{\frac{3x+1}{2}},\frac{\sqrt{3}(y+1)}{2}}\Bra{\textstyle{\frac{3x}{2},\frac{\sqrt{3}y}{2}}}\otimes\ket{0}\bra{0}\nonumber\\
 &+\Ket{\textstyle{\frac{3x-2}{2},\frac{\sqrt{3}y}{2}}}\Bra{\textstyle{\frac{3x}{2},\frac{\sqrt{3}y}{2}}}\otimes\ket{1}\bra{1}\nonumber\\
 &+\Ket{\textstyle{\frac{3x+1}{2},\frac{\sqrt{3}(y-1)}{2}}}\Bra{\textstyle{\frac{3x}{2},\frac{\sqrt{3}y}{2}}}\otimes\ket{2}\bra{2}\Biggr\},\\[3mm]
 S_2=\sum_{x,y\in\mathbb{Z}} & \Biggl\{\Ket{\textstyle{\frac{3x}{2},\frac{\sqrt{3}(y-1)}{2}}}\Bra{\textstyle{\frac{3x+1}{2},\frac{\sqrt{3}y}{2}}}\otimes\ket{0}\bra{0}\nonumber\\
 &+\Ket{\textstyle{\frac{3x+3}{2},\frac{\sqrt{3}y}{2}}}\Bra{\textstyle{\frac{3x+1}{2},\frac{\sqrt{3}y}{2}}}\otimes\ket{1}\bra{1}\nonumber\\
 &+\Ket{\textstyle{\frac{3x}{2},\frac{\sqrt{3}(y+1)}{2}}}\Bra{\textstyle{\frac{3x+1}{2},\frac{\sqrt{3}y}{2}}}\otimes\ket{2}\bra{2}\Biggr\},
\end{align}
and
\begin{align}
 C=\sum_{\chi,\upsilon\in\mathbb{H}}\ket{\chi,\upsilon}\bra{\chi,\upsilon}\otimes &
 \Biggl\{-\frac{1+c}{2}\ket{0}\bra{0}+\frac{s}{\sqrt{2}}\ket{0}\bra{1}+\frac{1-c}{2}\ket{0}\bra{2}\nonumber\\
 &+\frac{s}{\sqrt{2}}\ket{1}\bra{0}+c\ket{1}\bra{1}+\frac{s}{\sqrt{2}}\ket{1}\bra{2}\nonumber\\
 &+\frac{1-c}{2}\ket{2}\bra{0}+\frac{s}{\sqrt{2}}\ket{2}\bra{1}-\frac{1+c}{2}\ket{2}\bra{2}\Biggr\}\nonumber\\[3mm]
 =\sum_{\chi,\upsilon\in\mathbb{H}}\ket{\chi,\upsilon}\bra{\chi,\upsilon}\otimes &
  \begin{bmatrix}
   -\frac{1+c}{2}& \frac{s}{\sqrt{2}}& ~~\frac{1-c}{2}\\[2mm]
   ~~\frac{s}{\sqrt{2}}& c& ~~\frac{s}{\sqrt{2}}\\[2mm]
   ~~\frac{1-c}{2}& \frac{s}{\sqrt{2}}& -\frac{1+c}{2}
  \end{bmatrix},
\end{align}
with $c=\cos\theta, s=\sin\theta\, (\theta\in [0,2\pi))$.
Since the behavior of the walker is obvious, we will not treat $\theta=0,\pi$.
When we set $c=-\frac{1}{3}, s=\frac{2\sqrt{2}}{3}$, the coin-flip operator $C$ becomes a Grover coin
\begin{equation}
 C=\sum_{\chi,\upsilon\in\mathbb{H}}\ket{\chi,\upsilon}\bra{\chi,\upsilon}\otimes
  \begin{bmatrix}
   -\frac{1}{3}& \frac{2}{3}& \frac{2}{3}\\[2mm]
   \frac{2}{3}& -\frac{1}{3}& \frac{2}{3}\\[2mm]
   \frac{2}{3}& \frac{2}{3}& -\frac{1}{3}
  \end{bmatrix}.
\end{equation}
Assuming $\braket{\Psi_0|\Psi_0}=1$, the probability that the walker is observed at position $(\chi,\upsilon)\in\mathbb{H}$, is defined by
\begin{align}
 \mathbb{P}\left[(X_t,Y_t)=(\chi,\upsilon)\right]=\bra{\Psi_t}\left(\ket{\chi,\upsilon}\bra{\chi,\upsilon}\otimes\sum_{j=0}^{2} \ket{j}\bra{j}\right)\ket{\Psi_t},
\end{align}
where $(X_t,Y_t)$ denotes the position of the walker at time $t$.
Finally we set an initial condition
\begin{equation}
 \ket {\Psi_0}=\ket{0,0}\otimes\left(\alpha\ket{0} + \beta\ket{1} + \gamma\ket{2}\right),\label{eq:initial_state}
\end{equation}
for $\alpha,\beta,\gamma\in\mathbb{C}$ such that $|\alpha|^2+|\beta|^2+|\gamma|^2=1$, where $\mathbb{C}$ means the set of complex numbers.
This means the walker starts from the origin, and if it takes to be the initial state Eq.~(\ref{eq:initial_state}), the walk on a hexagonal lattice is obviously realized by the position-shift operators $S_1,S_2$.

The position-shift operator $S_2$ (resp. $S_1$) does not work to any whole system such that $\ket{\Psi}\in\mathcal{H}_{p_1}$ (resp. $\in\mathcal{H}_{p_2}$), and since $C\ket{\Psi}\in\mathcal{H}_{p_1}$ (resp. $\in\mathcal{H}_{p_2}$), we see $(S_1+S_2)C\ket{\Psi}=S_1C\ket{\Psi}$ (resp. $=S_2C\ket{\Psi}$).
Combining this fact and the initial state Eq.~(\ref{eq:initial_state}), we should note that the whole system becomes $\ket{\Psi_{2t}}=\left(S_2CS_1C\right)^t\ket{\Psi_0}\in\mathcal{H}_{p_1}$ and $\ket{\Psi_{2t+1}}=S_1C\left(S_2CS_1C\right)^t\ket{\Psi_0}\in\mathcal{H}_{p_2}$.
The walker, therefore, is not observed on $\mathbb{H}_1$ at time $2t+1$.
Figure~\ref{fig:lattice} explains the lattice $\mathbb{H}=\mathbb{H}_1\cup\mathbb{H}_2$ and the vertices where the walker arrives according to the position-shift operators $S_1, S_2$. 

\begin{figure}[h]
 \begin{center}
  \begin{minipage}{60mm}
   \includegraphics[scale=0.5]{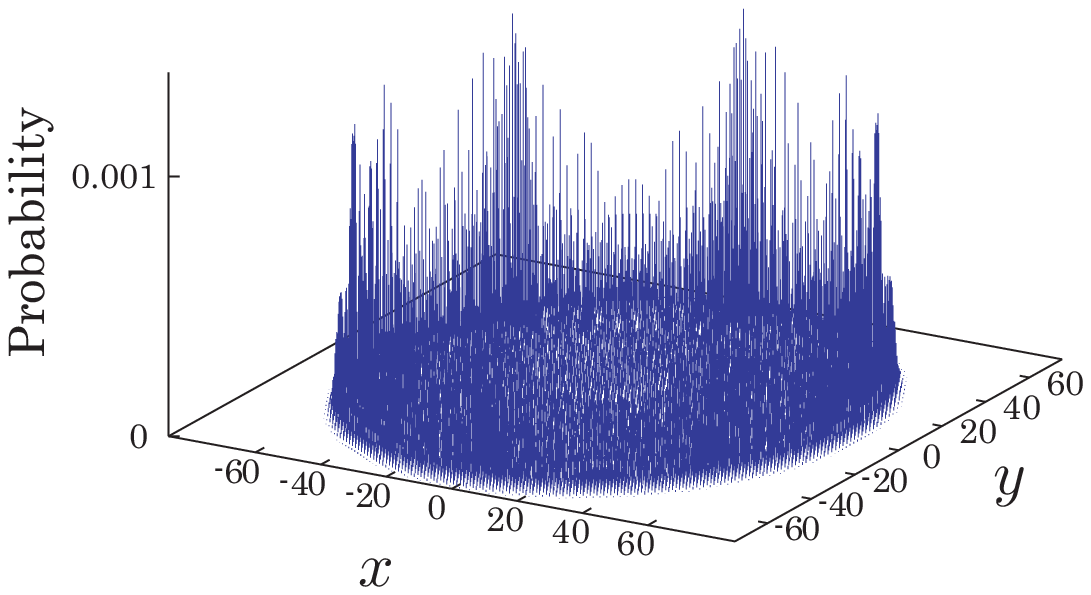}\\
   {(a) $\alpha=\beta=\gamma=\frac{1}{\sqrt{3}}$}
  \end{minipage}
  \begin{minipage}{60mm}
   \includegraphics[scale=0.5]{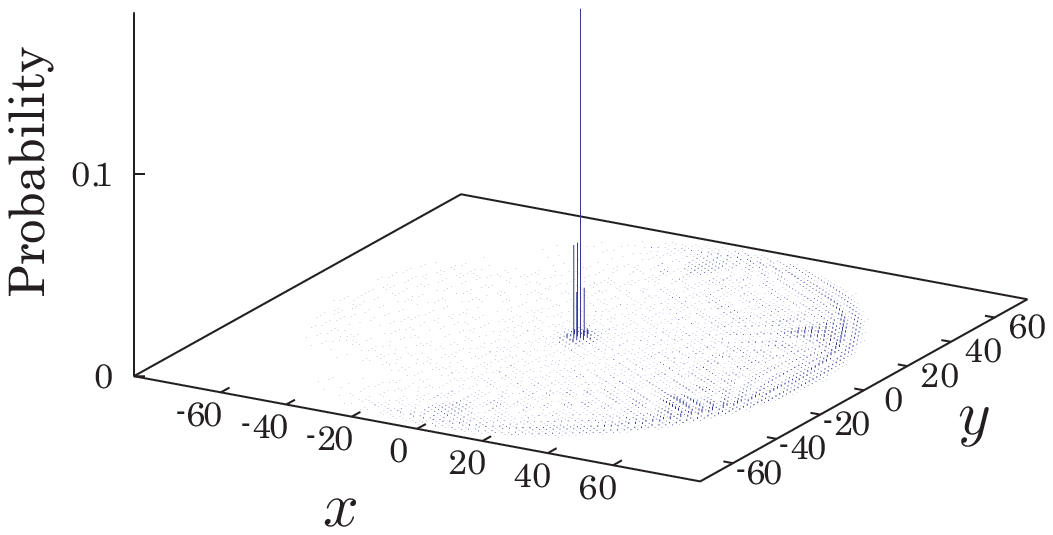}\\
   {(b) $\alpha=\gamma=0,\, \beta=1$}
   \end{minipage}
  \fcaption{Probability distribution at time $t=100$ ($c=-\frac{1}{3}, s=\frac{2\sqrt{2}}{3}$)}
  \label{fig:distribution} 
 \end{center}
\end{figure}

\noindent Moreover, as we can see from Fig.~\ref{fig:distribution}, the probability distribution sharply depends on the initial condition.

\section{Limit law of the return probability}
\label{sec:return_probability}

In this section we show a long-time limit law of the return probability.
The return probability is studied as one of the interesting subjects in the field of classical random walks because it is involved with the recurrence.
Similarly, the return probability of quantum walks is also interesting with the recurrence, compared with the classical ones.
Regarding quantum walks on the 2-dimensional square lattice, the relations between the return probability and the recurrence were discussed for some special walks~\cite{vStefavnakKissJex2008,vStefavnakJexKiss2008}.
As we already defined the model in the previous section, the walker starts from the origin.
So, the return probability means the probability that the walker can be observed at the origin (i.e. $\mathbb{P}[(X_t,Y_t)=(0,0)]$).
Since $\ket{\psi_{2t+1}(0,0)}={}^T[0,0,0]$ (i.e. $\mathbb{P}[(X_{2t+1},Y_{2t+1})=(0,0)]=0$) because of $\ket{\Psi_{2t+1}(0,0)}\in\mathcal{H}_{p_2}$, we should focus on the probability at the origin at time $2t$, where $T$ means the transposed operator.
For the return probability, one can get the following limit.
\bigskip

\begin{thm}
If the walker starts from the origin, we have a long-time limit of the return probability
 \begin{align}
  \lim_{t\to\infty}\mathbb{P}\left[(X_{2t},Y_{2t})=(0,0)\right]=&\Biggl|\left(\frac{1}{2}-\frac{A(\theta)}{\pi}\right)\alpha
    -\frac{\sqrt{2}sA(\theta)}{\pi (1-c)}\beta
    +\left\{\frac{(3+c)A(\theta)}{\pi(1-c)}-\frac{1}{2}\right\}\gamma\Biggr|^2\nonumber\\
  &+\Biggl|\frac{\sqrt{2}A(\theta)}{\pi (1-c)}\left\{s\alpha-\sqrt{2}(1-c)\beta+s\gamma\right\}\Biggr|^2\nonumber\\
  &+\Biggl|\left\{\frac{(3+c)A(\theta)}{\pi(1-c)}-\frac{1}{2}\right\}\alpha
	  -\frac{\sqrt{2}sA(\theta)}{\pi (1-c)}\beta
	  +\left(\frac{1}{2}-\frac{A(\theta)}{\pi}\right)\gamma\Biggr|^2,\label{eq:limit}
 \end{align}
 where $A(\theta)=\arcsin\left(\frac{1-\cos\theta}{3+\cos\theta}\right)=\arcsin\left(\frac{1-c}{3+c}\right)$.
 \label{th:return}
\end{thm}
\bigskip

\begin{figure}[h]
 \begin{center}
  \begin{minipage}{60mm}
   \includegraphics[scale=0.4]{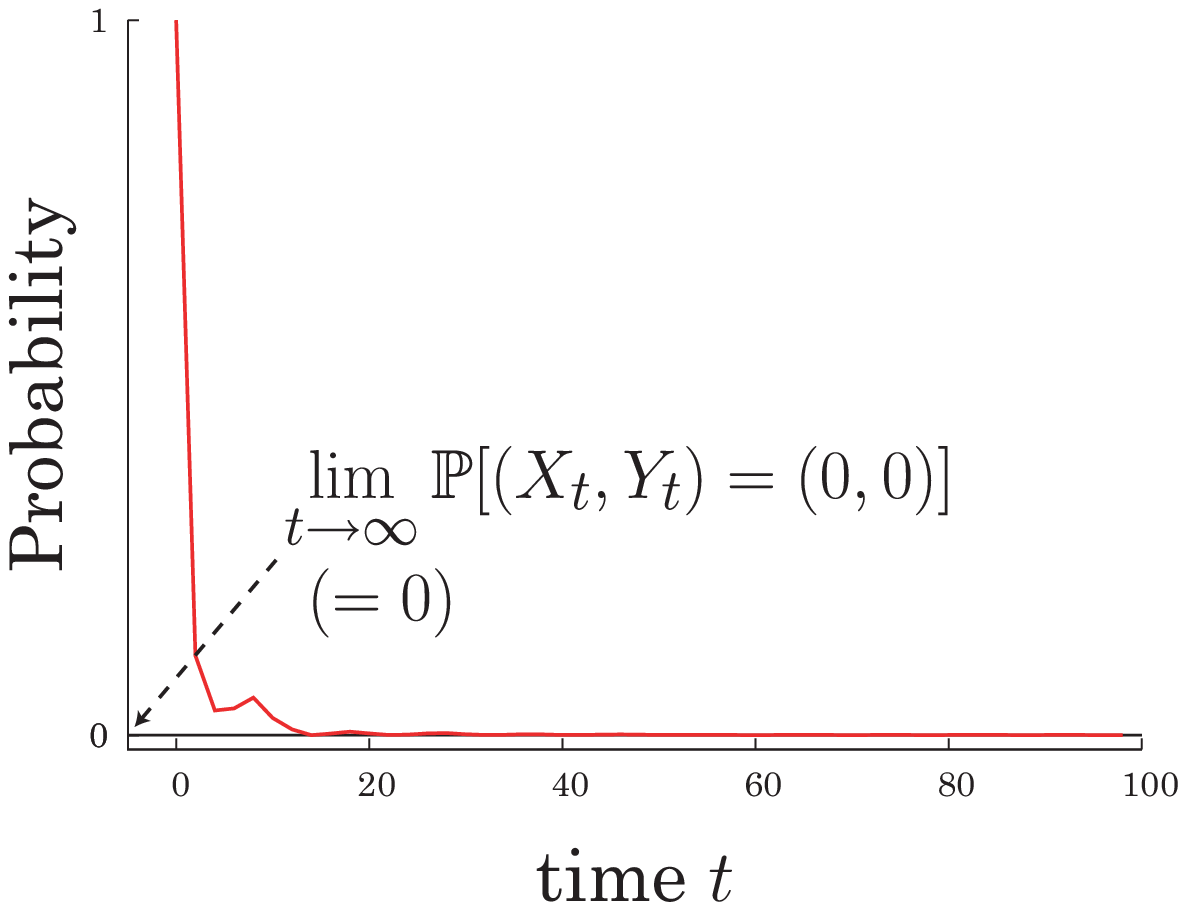}\\
   {(a) $\alpha=\beta=\gamma=\frac{1}{\sqrt{3}}$}
  \end{minipage}
  \begin{minipage}{60mm}
   \includegraphics[scale=0.4]{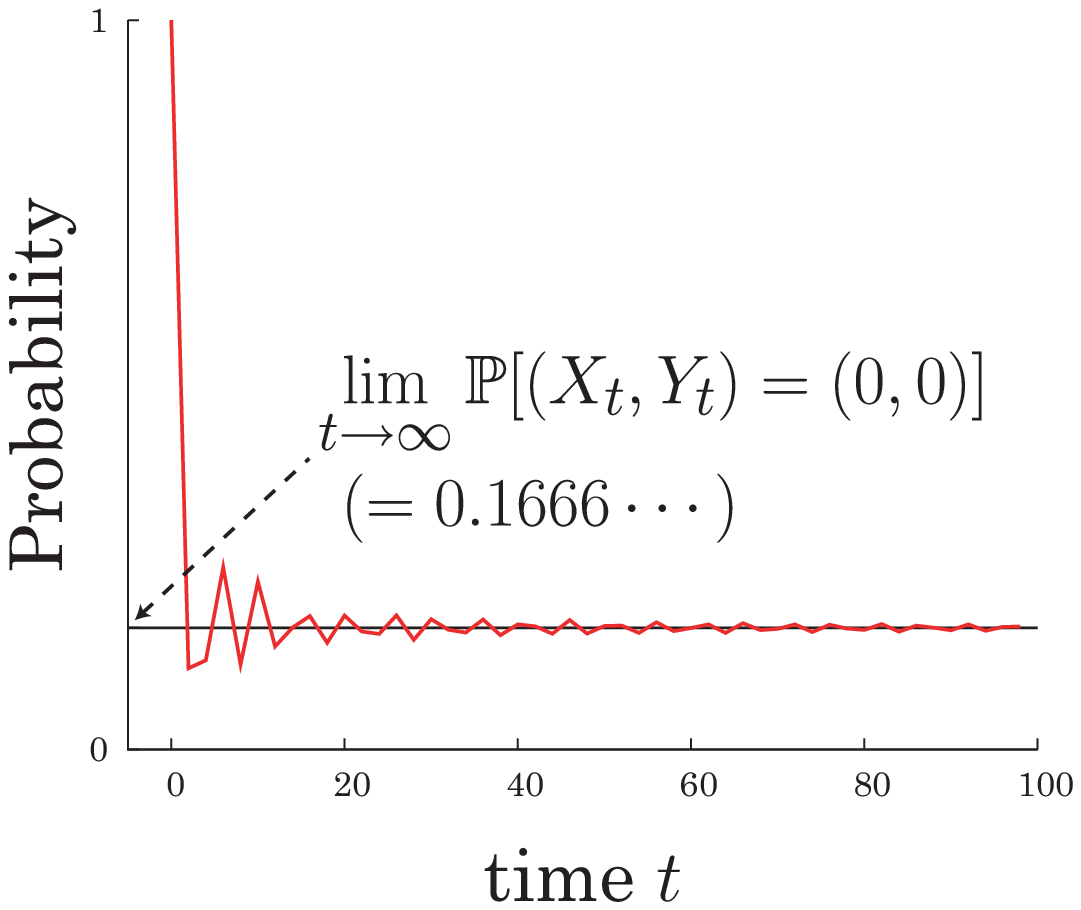}\\
   {(b) $\alpha=\gamma=0,\, \beta=1$}
   \end{minipage}
  \fcaption{The probability at the origin at even time ($c=-\frac{1}{3}, s=\frac{2\sqrt{2}}{3}$)}
  \label{fig:localization}
 \end{center}
\end{figure}

\noindent Figure~\ref{fig:localization} shows how the return probability $\mathbb{P}[(X_t,Y_t)=(0,0)]$ numerically converges to the limit Eq.~(\ref{eq:limit}).

\begin{proof}{%
We define a transform $\ket{\hat{\Psi}_t(a,b)}\in\mathbb{C}^3\,(a,b\in [-\pi,\pi))$ of the walk at time $t$ as
\begin{equation}
 \ket{\hat{\Psi}_t(a,b)}=\sum_{x,y\in\mathbb{Z}}e^{-i(ax+by)}\Ket{\psi_t\left(\textstyle{\frac{3x}{2},\frac{\sqrt{3}y}{2}}\right)} + \sum_{x,y\in\mathbb{Z}}e^{-i(ax+by)}\Ket{\psi_t\left(\textstyle{\frac{3x+1}{2},\frac{\sqrt{3}y}{2}}\right)},
\end{equation}
If the walker starts from the origin, the transform is reduced to a simpler form, according to even or odd time:
\begin{align}
 \ket{\hat{\Psi}_{2t}(a,b)}=&\sum_{x,y\in\mathbb{Z}}e^{-i(ax+by)}\Ket{\psi_{2t}\left(\textstyle{\frac{3x}{2},\frac{\sqrt{3}y}{2}}\right)},\\
 \ket{\hat{\Psi}_{2t+1}(a,b)}=&\sum_{x,y\in\mathbb{Z}}e^{-i(ax+by)}\Ket{\psi_{2t+1}\left(\textstyle{\frac{3x+1}{2},\frac{\sqrt{3}y}{2}}\right)}.
\end{align}
That's because we have $\ket{\psi_{2t}(\chi,\upsilon)}={}^T[0,0,0]$ at position $(\chi,\upsilon)\in\mathbb{H}_2$ and $\ket{\psi_{2t+1}(\chi,\upsilon)}={}^T[0,0,0]$ at position $(\chi,\upsilon)\in\mathbb{H}_1$, for any $t\in\left\{0,1,2,\ldots\right\}$.
The amplitude at the position where the walker can be observed, is extracted from a transform
\begin{align}
 \Ket{\psi_{2t}\left(\textstyle{\frac{3x}{2},\frac{\sqrt{3}y}{2}}\right)}=&\int_{-\pi}^\pi\frac{da}{2\pi} \int_{-\pi}^\pi\frac{db}{2\pi}\,\, e^{i(ax+by)}\ket{\hat{\Psi}_{2t}(a,b)},\\
 \Ket{\psi_{2t+1}\left(\textstyle{\frac{3x+1}{2},\frac{\sqrt{3}y}{2}}\right)}=&\int_{-\pi}^\pi\frac{da}{2\pi} \int_{-\pi}^\pi\frac{db}{2\pi}\,\, e^{i(ax+by)}\ket{\hat{\Psi}_{2t+1}(a,b)}.
\end{align}
As we can see from this inverse transform, the amplitude is obtained just like the inverse Fourier transform.
We, hence, call the transform $\ket{\hat\Psi_t(a,b)}$ the Fourier transform in this paper.  
Equation~(\ref{eq:time_evo}) leads us to the time-evolution of the Fourier transform
\begin{align}
 \ket{\hat\Psi_{2t+1}(a,b)}=&R(a,b)\tilde{C}\ket{\hat\Psi_{2t}(a,b)},\label{eq:rec1}\\
 \ket{\hat\Psi_{2t+2}(a,b)}=&R(-a,-b)\tilde{C}\ket{\hat\Psi_{2t+1}(a,b)}\label{eq:rec2},
\end{align}
where the matrix $R(a,b)$ is a $3\times 3$ diagonal unitary matrix
\begin{align}
 R(a,b)=&e^{-ib}\ket{0}\bra{0}+e^{ia}\ket{1}\bra{1}+e^{ib}\ket{2}\bra{2}\nonumber\\[2mm]
 =&\begin{bmatrix}
    e^{-ib} & 0 & 0 \\
    0 & e^{ia} & 0\\
    0 & 0 & e^{ib}
   \end{bmatrix},
\end{align}
and
\begin{equation}
 \tilde{C}=\begin{bmatrix}
	    -\frac{1+c}{2}& \frac{s}{\sqrt{2}}& ~~\frac{1-c}{2}\\[2mm]
	    ~~\frac{s}{\sqrt{2}}& c& ~~\frac{s}{\sqrt{2}}\\[2mm]
	    ~~\frac{1-c}{2}& \frac{s}{\sqrt{2}}& -\frac{1+c}{2}
	   \end{bmatrix}.
\end{equation}
From Eqs.~(\ref{eq:rec1}) and (\ref{eq:rec2}), the Fourier transform at time $2t$ turns out to be
\begin{equation}
 \ket{\hat{\Psi}_{2t}(a,b)}=\left\{R(-a,-b)\tilde{C}R(a,b)\tilde{C}\right\}^t\ket{\hat{\Psi}_0(a,b)}.
\end{equation}
We express the eigenvalues $\lambda_j(a,b)\,(j=1,2,3)$ of the unitary matrix $R(-a,-b)\tilde{C}R(a,b)\tilde{C}$ as follows:
\begin{equation}
 \lambda_j(a,b)=e^{i\nu_j(a,b)}\quad(j=1,2,3),
\end{equation}
with
\begin{equation}
 \left\{\begin{array}{ll}
   \nu_1(a,b)=&0,\\[1mm]
 \nu_2(a,b)=&\arccos\left\{c^2-\frac{1}{2}(1-c)^2\sin^2 b+s^2\cos a\cos b\right\},\\[1mm]
 \nu_3(a,b)=&2\pi-\arccos\left\{c^2-\frac{1}{2}(1-c)^2\sin^2 b+s^2\cos a\cos b\right\}.
	\end{array}\right.
\end{equation}

Here, for $x,y\in\mathbb{Z}$, we put
\begin{equation}
 g(x,y)=\frac{1}{4\pi^2}\int_{-\pi}^\pi da \int_{-\pi}^\pi db\,\,\frac{e^{i(ax+by)}}{2 s^2(1-\cos a \cos b)+(1-c)^2\sin^2 b}.\label{eq:definition_g}
\end{equation}
By using the residue theorem with regard to the parameter $a$ in Eq.~(\ref{eq:definition_g}), we get an integral representation
\begin{equation}
 g(x,y)=\int_0^\pi \frac{\cos(b|y|) \left\{2s^2+(1-c)^2\sin^2 b-(1-c)\sin b\sqrt{(3+c)^2-(1-c)^2\cos^2 b}\right\}^{|x|}}{\pi(1-c)\sin b(2s^2\cos b)^{|x|}\sqrt{(3+c)^2-(1-c)^2\cos^2 b}}\,db.
\end{equation}
Again, for $x,y\in\mathbb{Z}$, one can obtain a long-time asymptotic behavior of the amplitude at position $\left(\frac{3x}{2},\frac{\sqrt{3}y}{2}\right)\in\mathbb{H}_1$,
\begin{align}
 &\Ket{\psi_{2t}\left( \textstyle{\frac{3x}{2},\frac{\sqrt{3}y}{2}}\right)}\nonumber\\
 = & \int_{-\pi}^\pi \frac{da}{2\pi} \int_{-\pi}^\pi \frac{db}{2\pi}\,\,\sum_{j=1}^3 e^{i(ax+by)}\lambda_j(a,b)^t \braket{v_j(a,b)|\hat\Psi_0(a,b)}\ket{v_j(a,b)}\nonumber\\
 \sim & \int_{-\pi}^\pi \frac{da}{2\pi} \int_{-\pi}^\pi \frac{db}{2\pi}\,\,e^{i(ax+by)}\braket{v_1(a,b)|\hat\Psi_0(a,b)}\ket{v_1(a,b)}\quad (t\to\infty)\nonumber\\[3mm]
 = & \begin{bmatrix}
      -\frac{s}{2}\left\{ W_1(\alpha,\gamma) G(x,y,1,-1) + W_2(\alpha,\beta) G(x+1,y-1,1,-1) + W_2(\gamma,\beta) G(x,y+2,-1,1) \right\}\\[2mm]
      -\frac{\sqrt{2}}{4}(1-c)\left\{ W_1(\alpha,\gamma) G(x-1,y+1,0,2) + W_2(\alpha,\beta) G(x,y,0,2) + W_2(\gamma,\beta) G(x,y,0,-2) \right\}\\[2mm]
      \frac{s}{2}\left\{ W_1(\alpha,\gamma) G(x,y,1,1) + W_2(\alpha,\beta) G(x+1,y-1,1,1) + W_2(\gamma,\beta) G(x,y,-1,-1) \right\}
     \end{bmatrix},\label{eq:asymptotic_behavior}
\end{align}
where $\ket{v_j(a,b)}\,(j=1,2,3)$ are normalized eigenvectors corresponding to the eigenvalues $\lambda_j(a,b)$ and we have put
\begin{align}
 G(x,y,x_1,y_1)= & g(x,y)-g(x-x_1,y-y_1),\\
 W_1(z_1,z_2)= & -sz_1+sz_2,\\
 W_2(z_1,z_2)= & sz_1-\frac{\sqrt{2}}{2}(1-c)z_2.
\end{align}
The asymptotic symbol $h_1(t)\sim h_2(t)\,(t\to\infty)$ means $\lim_{t\to\infty}h_1(t)/h_2(t)=1$.
The Riemann-Lebesgue lemma have been used in Eq.~(\ref{eq:asymptotic_behavior}).
Straightforwardly computing the long-time asymptotic behavior of the amplitude at the origin, we have
 \begin{align}
  \ket{\psi_{2t}(0,0)}\sim&
  \begin{bmatrix}
   \left(\frac{1}{2}-\frac{A(\theta)}{\pi}\right)\alpha
   -\frac{\sqrt{2}sA(\theta)}{\pi (1-c)}\beta
   +\left\{\frac{(3+c)A(\theta)}{\pi(1-c)}-\frac{1}{2}\right\}\gamma\\[3mm]
   -\frac{\sqrt{2}A(\theta)}{\pi (1-c)}\left\{s\alpha-\sqrt{2}(1-c)\beta+s\gamma\right\}\\[3mm]
   \left\{\frac{(3+c)A(\theta)}{\pi(1-c)}-\frac{1}{2}\right\}\alpha
   -\frac{\sqrt{2}sA(\theta)}{\pi (1-c)}\beta
   +\left(\frac{1}{2}-\frac{A(\theta)}{\pi}\right)\gamma
  \end{bmatrix}\quad(t\to\infty).
 \end{align}
This also gives the long-time limit of the return probability.
}
\end{proof}

\section{Discussion and summary}
The return probability of symmetric simple random walks on a lattice strongly depends on the structure of the lattice, while that of quantum walks depends on their coin-flip operator.
It have not been cleared yet how the return probability of quantum walkers is determined by the structure of a lattice.
We treated a quantum walk on a hexagonal lattice and its coin-flip operator was given so that it included a Grover coin.
Since the walker started from the origin, the return probability was simply defined by the probability that the walker can be observed at the origin.  
As a result, we derived a long-time limit theorem about the return probability.
Also, Theorem~\ref{th:return} gives a condition that either localization or delocalization at the origin occurs.
When we get $\limsup_{t\to\infty}\mathbb{P}[(X_t,Y_t)=(\chi,\upsilon)]=0$ (resp. $>0$) for position $(\chi,\upsilon)\in\mathbb{H}$, let us say that delocalization (resp. localization) at position $(\chi,\upsilon)$ occurs.
From this definition, we find a condition of delocalization at the origin.
Delocalization at the origin is realized if and only if
\begin{equation}
 |\alpha|=\frac{\sqrt{1-c}}{2},\, \beta=\frac{\sqrt{2}(1+c)}{s}\alpha,\, \gamma=\alpha.\label{eq:delocalized}
\end{equation}
The condition of Figs.~\ref{fig:distribution}-(a) and \ref{fig:localization}-(a) satisfies Eq.~(\ref{eq:delocalized}) under the condition $c=-\frac{1}{3}, s=\frac{2\sqrt{2}}{3}$.

On the other hand, a limit theorem on a rescaled space by time $t$ can be also demonstrated.

\begin{thm}
There exists a continuous function $f:\mathbb{R}^2\longmapsto\mathbb{R}$ such that, for $x,y\in\mathbb{R}$, we have
\begin{equation}
 lim_{t\to\infty}\mathbb{P}\left(\frac{2X_t}{3t}\leq x,\,\frac{2Y_t}{\sqrt{3}t}\leq y\right)=\int_{-\infty}^x du \int_{-\infty}^y dv\,\,\Delta(\alpha,\beta,\gamma)\delta_o(u,v)+f(u,v),
\end{equation}
where $\delta_o(x,y)$ is the Dirac $\delta$-function at the origin and
\begin{align}
 \Delta(\alpha,\beta,\gamma)=&\left(\frac{1}{2}-\frac{A(\theta)}{\pi}\right)|\alpha|^2+\frac{2A(\theta)}{\pi}|\beta|^2+\left(\frac{1}{2}-\frac{A(\theta)}{\pi}\right)|\gamma|^2\nonumber\\
 &-\frac{2\sqrt{2}sA(\theta)}{\pi(1-c)}\Re\left\{(\alpha+\gamma)\overline{\beta}\right\}+\left\{\frac{2(3+c)A(\theta)}{\pi(1-c)}-1\right\}\Re(\alpha\overline{\gamma}).
\end{align}
We mean that $\mathbb{R}$ is the set of real numbers and $\Re(z)\,(z\in\mathbb{C})$ is the real part of the complex number $z$.
\end{thm}
\bigskip

\noindent We can also consider the value $\Delta(\alpha,\beta,\gamma)$ as a return probability on the rescaled space.
Although this theorem is obtained by using the Fourier analysis like the other past researches (e.g.~\cite{GrimmettJansonScudo2004,WatabeKobayashiKatoriKonno2008,MachidaChandrashekarKonnoBusch2013}), it is not completed because the continuous function $f(x,y)$ is not computed.
Note that we have $\Delta(\alpha,\beta,\gamma)=0$ for the initial state such that Eq.~(\ref{eq:delocalized}).
It would be one of the interesting future problems to compute the continuous part $f(x,y)$.

\nonumsection{Acknowledgements}
The author TM acknowledges support from the Japan Society for the Promotion of Science.
Also, he would like to thank F.~Alberto~Gr{\"u}nbaum and Luis Vel\'azquez for a useful discussion.
\bigskip


\end{document}